\documentclass[conference]{IEEEtran}
\IEEEoverridecommandlockouts

\usepackage{cite}
\usepackage{amsmath,amssymb,amsfonts}
\usepackage{algorithmic}
\usepackage{graphicx}
\usepackage{textcomp}
\usepackage{xcolor}
\usepackage{makecell}
\usepackage{booktabs}
\usepackage{tabularx}
\usepackage{multirow}
\usepackage{amsfonts,amssymb}
\usepackage{hyperref}
\usepackage{url}
\usepackage{caption}
\usepackage[table]{xcolor}
\usepackage{subcaption}
\definecolor{mycyan}{HTML}{57B5C3}
\definecolor{mycyan1}{HTML}{DCC7A9}
\definecolor{mycyan2}{HTML}{CBA67B}
\usepackage{marvosym}
\usepackage{textcomp}
\def\BibTeX{{\rm B\kern-.05em{\sc i\kern-.025em b}\kern-.08em
    T\kern-.1667em\lower.7ex\hbox{E}\kern-.125emX}}
\author{
\IEEEauthorblockN{
Hongqin Lyu\textsuperscript{1,2},
Yonghao Wang\textsuperscript{1,2},
Yunlin Du\textsuperscript{5}, 
Mingyu Shi\textsuperscript{4},
Zhiteng Chao\textsuperscript{1},
Wenxing Li\textsuperscript{1},\\
Tiancheng Wang\textsuperscript{1,2,3\dag}
and
Huawei Li\textsuperscript{1,2,3\dag}}
\IEEEauthorblockA{\textsuperscript{1}State Key Lab of Processors, Institute of Computing Technology, CAS, Beijing, China}
\IEEEauthorblockA{\textsuperscript{2}University of Chinese Academy of Sciences, Beijing, China}
\IEEEauthorblockA{\textsuperscript{3}CASTEST, Beijing, China}
\IEEEauthorblockA{\textsuperscript{4}School of Integrated Circuits, Nanjing University, Suzhou, China}
\IEEEauthorblockA{\textsuperscript{5}School of Information and Physical Sciences, University of Newcastle, Newcastle, Australia}
}
\begin{document}

\title{AssertGen: Enhancement of LLM-aided Assertion Generation through Cross-Layer Signal Bridging
\thanks{\textsuperscript{\dag}\ Corresponding author. \\
This paper has been accepted by the 34th IEEE Asian Test Symposium (ATS 2025), 
December 16--19, 2025, Tokyo, Japan.}}

\maketitle
\begin{abstract}
Assertion-based verification (ABV) serves as a crucial technique for ensuring that register-transfer level (RTL) designs adhere to their specifications. While Large Language Model (LLM) aided assertion generation approaches have recently achieved remarkable progress, existing methods are still unable to effectively identify the relationship between design specifications and RTL designs, which leads to the insufficiency of the generated assertions. To address this issue, we propose AssertGen, an assertion generation framework that automatically generates SystemVerilog assertions (SVA). AssertGen first extracts verification objectives from specifications using a chain-of-thought (CoT) reasoning strategy, then bridges corresponding signals between these objectives and the RTL code to construct a cross-layer signal chain, and finally generates SVAs based on the LLM. Experimental results demonstrate that AssertGen outperforms the existing state-of-the-art methods across several key metrics, such as pass rate of formal property verification (FPV), cone of influence (COI), proof core and mutation testing coverage.

\end{abstract}


\section{Introduction}
Functional verification plays a key role in the integrated circuit (IC) design process, which is essential to maintain the correctness and reliability of the design. Assertion-based verification (ABV) has emerged as an effective verification technique at register-transfer level (RTL), and been widely adopted by IC design projects. It has been reported that \cite{b1}, more than 70\% of ASIC design projects adopt ABV and the adoption percentage is still growing year by year, because ABV improves observability and controllability of designs, and accordingly, allows for dramatic reduction in verification time and debug efforts.

However, assertion generation typically relies on manual efforts, which is time-consuming, error-prone, and susceptible to omissions. With the growing scale and complexity of modern IC designs, the demand for automatic and high-quality assertion generation is becoming increasingly urgent.
Fortunately, recent advances in Large language model (LLM) have facilitated effective applications in various hardware verification tasks, such as stimuli generation \cite{b2,b3}, security verification \cite{b4,b5}, and formal verification \cite{b6,b7}.  The power of LLM has also promoted research in assertion generation of hardware designs.

Existing LLM-based assertion generation methods can be categorized into two categories according to their application stage: RTL-stage generation and Pre-RTL-stage generation. The RTL-stage assertion generation focuses on guiding LLM to analyze RTL codes and infer logical relations among signals to construct assertions \cite{b8}. However, since RTL designs may contain semantic errors or be incomplete, assertions generated from RTL designs, may inherit possible errors inside the code, potentially leading to mismatches with design specifications \cite{b9}.
\begin{figure}
\centering
\includegraphics[width=80mm]{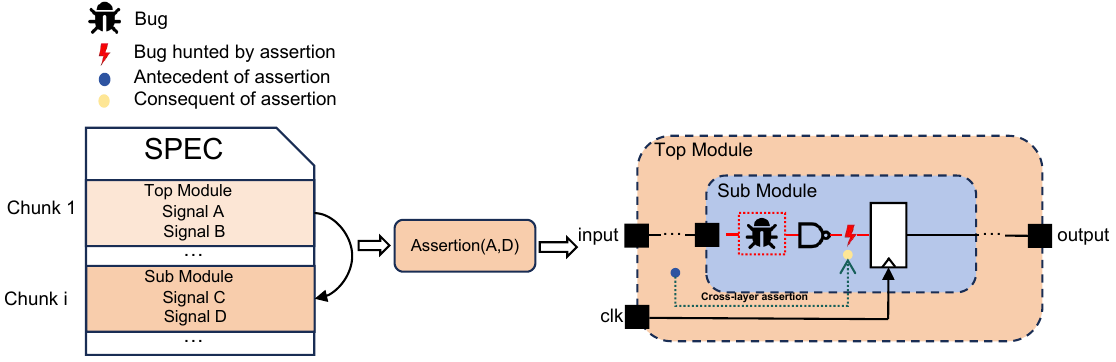}
\caption{Example: cross-layer assertion generation process and its role in bug detection}
\label{Cross-Layer Signal Bridging for Enhancing Assertion Generation}
\end{figure}
Pre-RTL-stage assertion generation focuses on the natural language specifications, aligning specifications with Verilog codes through semantic mapping to generate assertions\cite{b10}. Instead of relying solely on RTL codes, this method leverages information of design specifications to reducing the impact of deficiencies of RTL designs on assertion quality.

However, current methods that generate assertions only for the top-level module will lead to inaccurate bug detection when a problem occurs in a sub-level module. The root cause of this inadequacy lies in three gaps between the design specifications and the RTL code. Firstly, functional descriptions in specifications are often scattered across multi semantic levels and textual chunks, creating a structural gap that makes signal behavior extraction an inherent cross-layer challenge. Secondly, a semantic gap exists between behavioral descriptions and verification objectives\cite{b11}, which is mainly due to the fragmentation of information. Specifically, the composition logic of a complete and rigorous verification objectives is often scattered in multiple text chunks, which spans different semantic levels. Lastly, there also have a link gap from verification objectives to corresponding RTL code segments, which makes it challenging to fully connect verification objectives with related code segments.

To address these challenges, we propose a Pre-RTL framework—AssertGen—to bridge the above three gaps: incremental context extraction strategy for the structural gap, chain-of-thought (CoT) prompting strategy for the semantic gap, and cross-layer signal chain strategy for the link gap. Fig. 1 shows how AssertGen generates cross-layer assertions by combining signals (e.g., Signal A and Signal D) described across different layers in the specification.

The main contributions of this work are as follows:
\begin{figure*}[!t]
\centering
\includegraphics[width=135mm]{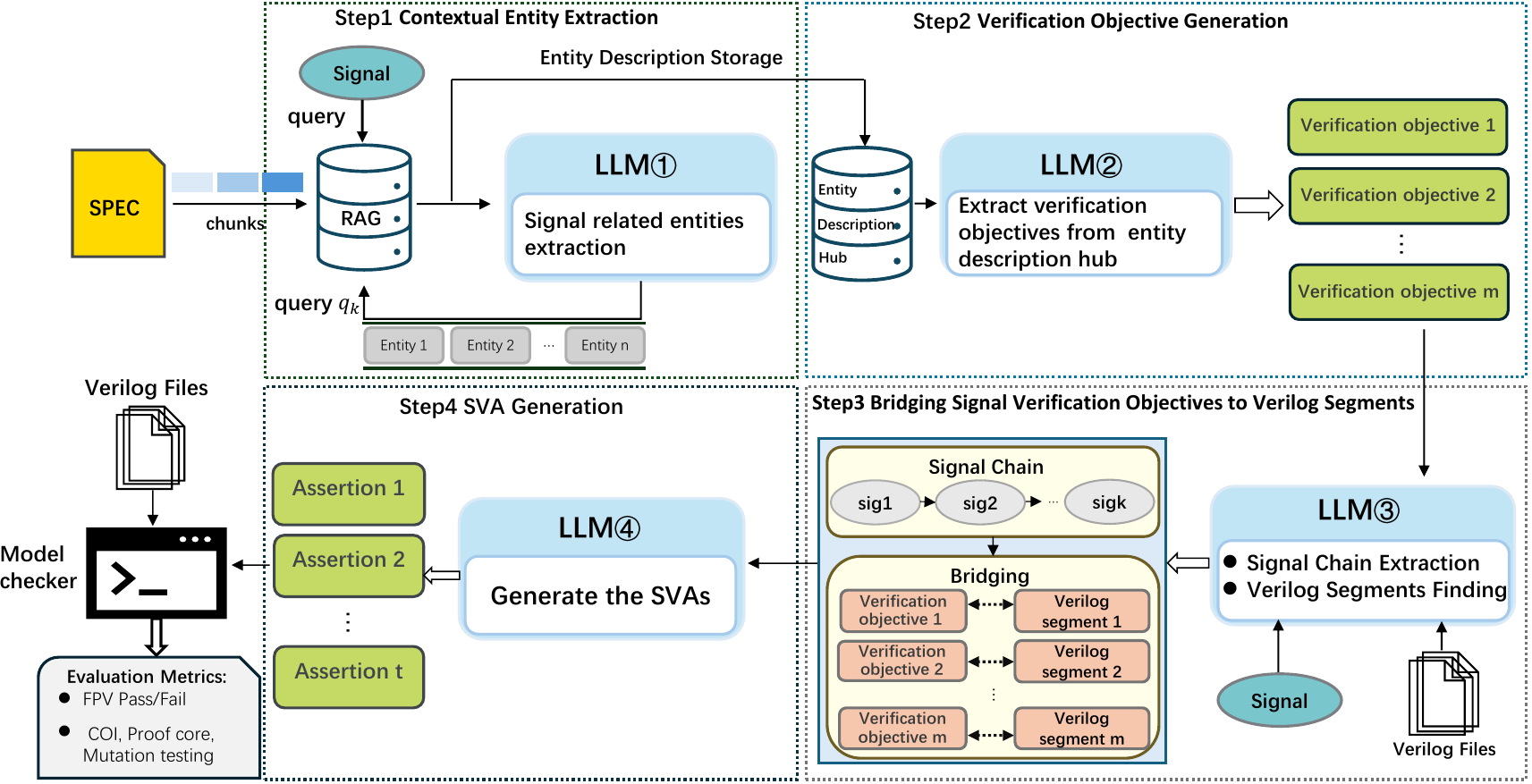}
\caption{AssertGen framework}
\label{AssertGenframework}
\end{figure*}
\begin{enumerate}[]
    \item Cross-layer signal chains are used to connect specifications and RTL designs, accurately mapping signal behavior to code and extracting detailed information to generate assertions.
    \item To address the problem of scattered signal behavior descriptions and complex semantics in specifications, we propose an incremental context extraction method combined with CoT prompt to generate more comprehensive verification objectives.
    \item We evaluate AssertGen on five open-source benchmark designs. Experimental results show that AssertGen surpasses current state-of-the-art methods in formal property verification pass rate, COI, proof-core coverage, and bug detection rate of mutation testing. 
\end{enumerate}
\section{Related work}
\subsection{Assertion generation based on NLP}
Early template-based assertion generation methods relied on static analysis of specifications, but their effectiveness was limited without HDL code \cite{b12}. To overcome this limitation, subsequent work focused on extracting properties directly from the RTL design itself \cite{b13}. However, this method was unreliable when the RTL design is flawed. To address this issue, a later approach was proposed  to generate assertions directly from the specification using subtree analysis, ensuring the assertions to reflect the intended design requirements \cite{b14}.

\subsection{Assertion generation based on LLM}
Leveraging LLMs is a promising direction for assertion generation, as their strong semantic comprehension addresses the shortcomings of traditional NLP methods.

Building on their known natural language processing capabilities, researchers are increasingly applying LLMs to hardware assertion generation. Early work first explored generating hardware security assertions \cite{b15}. Subsequent research has developed frameworks for dynamic simulation \cite{b8} and used simulation log feedback loops for iterative refinement \cite{b16}. To manage hardware's inherent complexity, strategies have been proposed such as assertion mining based on deep-level design analysis \cite{b17} and decomposing complex tasks into simpler subtasks \cite{b10} to improve assertion quality and quantity.
\section{AssertGen Framework}
\subsection{Overview}
To enhance the efficiency of assertion generation, we propose AssertGen, which aims to automatically generate correct assertions by cross-level information bridging and behavioral modeling, as shown in Fig. \ref{AssertGenframework}. The framework achieves its functionality through a four-step process, detailed as follows:
\begin{enumerate}[]
    \item \textbf{Contextual Entity Extraction:} An incremental strategy extracts signal behaviors and related entities from the specification to support verification objectives generation.
    \item \textbf{Verification Objective Generation:} By leveraging a prompt mechanism with CoT reasoning, verification objectives are generated with cross-signal behavior.
    \item \textbf{Bridging Verification Objectives to Verilog Segments:} A hierarchical mapping strategy using cross-layer signal chains is proposed to link verification objectives to their corresponding Verilog code.
    \item \textbf{SVA Generation:} Structured prompts are designed to alleviate inconsistent formatting in content generation by LLMs.
\end{enumerate}

\subsection{\textbf{Step 1: Contextual Entity Extraction}}
\begin{figure}
\centering
\includegraphics[width=55mm]{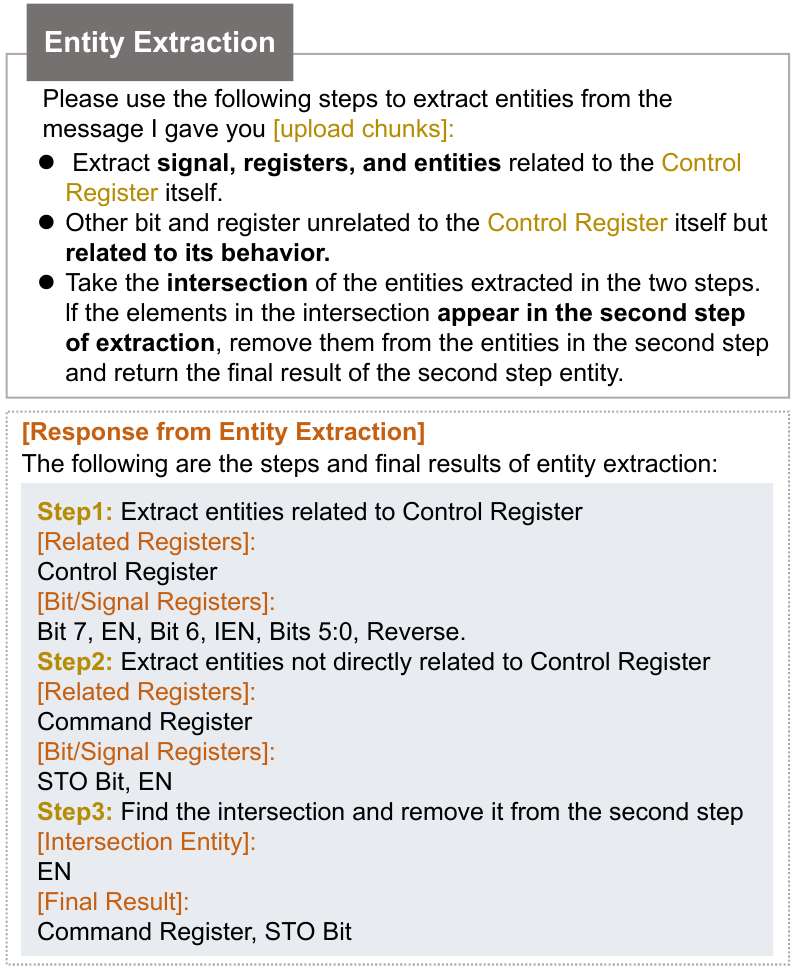}
\caption{Prompt and response of entity extraction}
\label{Entity Extraction prompt and response}
\end{figure}

In IC designs, signal behaviors often emerge from interactions among multiple signals, making their descriptions implicitly embedded in cross-signal contexts. To extract these semantic relationships, we propose to extract contextual entities from the specification, where entities refer to key elements such as signals, bits, or registers that convey behavioral meaning in the specification.

To effectively extract descriptions containing entities, we introduce a retrieval-augmented generation (RAG) mechanism. The core principle of RAG is to select the most relevant subset from a preprocessed collection of specification chunks (while chunks are a series of small, more easily retrievable text blocks) based on relevance computations with respect to a user query $q \in \mathcal{Q}$. The query follows a uniform template “\emph{What is the description of \{signal\}?},” where the initial signal used in the template to start Step 1 is arbitrarily selected from the specification at the framework’s first invocation.

Although RAG effectively yields signal-relevant descriptions, cross-signal information extraction remains challenging due to vague semantic cues. Our three-stage prompt strategy addresses this by first extracting entities that directly describe the target signal $E_{\text{target}}$, and then identifying all associated external entities$E_{\text{context}}$. To isolate purely contextual information, we then remove any entities that overlap with the target set, yielding $E_{\text{final}} = E_{\text{context}} - (E_{\text{target}} \cap E_{\text{context}})$. This method yields a set of highly-relevant contextual "probes," which serve as the basis for subsequent cross-layer information retrieval and for integrating related descriptions scattered across different text chunks.
Figure \ref{Entity Extraction prompt and response} shows the LLM-generated response for the Control Register of an I\textsuperscript{2}C design\cite{b18}.

After initializing the external entity set $E_{\text{final}}$ from the first query, the system iteratively removes each entity $e_k \in E_{\text{final}}$ which serves as a new query signal to construct the corresponding query $q_k$, and performs RAG-based retrieval over the vector database $\mathcal{H}$. This iterative expansion continues until no new entities are produced, i.e., $E_{\text{final}} = \varnothing$, marking the convergence of the entity set.  Meanwhile, all retrieved chunks $c_i$ from $\mathcal{H}$ are persistently stored to ensure that subsequent stages can effectively utilize the accumulated contextual information.
\subsection{\textbf{Step 2: Verification Objective Generation}}
\begin{figure}
\centering
\includegraphics[width=55mm]{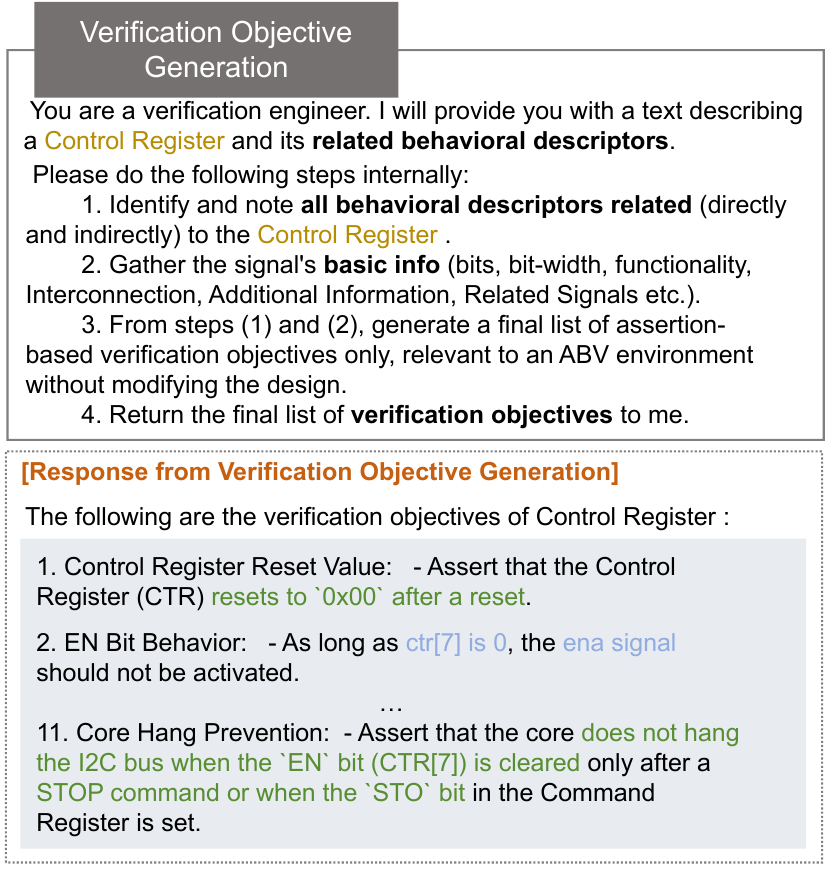}
\caption{Prompt and response of verification objective generation}
\label{Verification Features Generation prompt and response}
\end{figure}
In practical hardware development, engineers typically adopt a top-down methodology to modularize circuit design, thereby reducing inter-module coupling and simplifying integration and debugging. However, this methodology often lacks explicit bridging of signal behaviors across layers in the specification. To address this limitation, verification engineers abstract signal descriptions and contextual entities into verification objectives, enabling precise cross-layer bug detection and module-level attribution.

Building upon the advantages of verification objective abstraction, we adopt a similar strategy within the AssertGen framework. Specifically, we use the target signal $s$ and the retrieved specification chunks $\mathcal{C}'$ as input to the LLM to construct the verification objectives $v$:
\begin{equation}
    v = f_{\text{LLM}}(s, \mathcal{C}')
\end{equation}
To ensure the LLM can effectively integrate the graphical descriptions, structural features, and implicit temporal dependencies scattered across different semantic levels, we designed and applied a structured prompt as shown in Fig. \ref{Verification Features Generation prompt and response}, where the blue-highlighted parts of the verification objective includes both the top-level module and sub-level module. This prompt guides the model to generate a comprehensive and precise verification objectives set $\mathcal{V} = \{v_1, v_2, \dots, v_n\}$.
\subsection{\textbf{Step 3: Bridging Signal Verification Objectives to Verilog Segments}}
Requirement traceability, which provides bidirectional mapping between specifications, RTL designs, and verification objectives, is critical for ensuring every requirement is fully implemented and verified. However, the hierarchical nature of SoC designs and cross-layer signal propagation make this process highly complex.

To address this challenge, we propose a two-stage structured reverse bridging approach comprising: (1) Signal Chain Extraction, and (2) Bridging Verification objectives to Verilog Segments. This approach incrementally bridges each signal in a verification intent to its definition in the RTL hierarchy. 
\begin{figure}
\centering
\includegraphics[width=55mm]{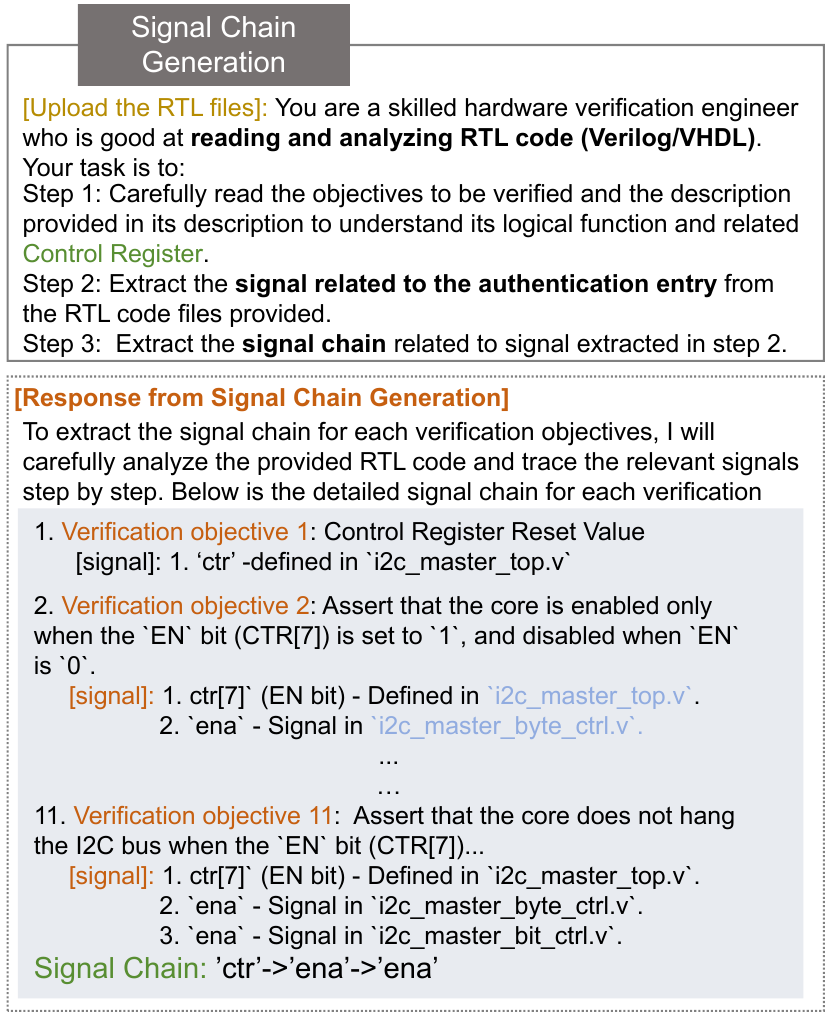}
\caption{Prompt and response of signal chain}
\label{Prompt and response of signal chain}
\end{figure}

\textbf{Signal Chain Extraction:} In hierarchical RTL designs, signals often form complex cross-layer propagation paths through various modules. Traditional language models struggle to accurately capture these structural dependencies. To this end, we define the concept of signal chain ($\mathcal{SC}$), as detailed below:

\textbf{Definition 1} \textbf{(Signal Chain).} Given a signal $s_{\text{0}} \in \text{Module}_0$, where $\text{Module}_0$ denotes the highest-level module in the design hierarchy where $s_0$ appears, and $s_{0}$ is included in the verification objectives, its cross-layer propagation path is represented as an ordered sequence:
\begin{equation}
    \mathcal{SC}(s_{\text{0}}) = \{s_{\text{0}}, s_1, \dots, s_n\}, \quad s_i \in \text{Module}_i
\end{equation}
where any adjacent signal pair satisfies the semantic propagation constraint:
\begin{equation}
    \forall i \in [1, n], \quad s_{i-1} \xrightarrow{\text{semantic}} s_{i}
\end{equation}
That is, each signal $s_i$ is semantically derived from its predecessor $s_{i-1}$.

Based on this definition, we design a recursive method grounded in CoT Prompting, which guides the LLM to construct signal dependencies across layers. This serves as a structural prior for the subsequent bridging phase, as illustrated in Fig. \ref{Prompt and response of signal chain}.

\begin{figure}
\centering
\includegraphics[width=55mm]{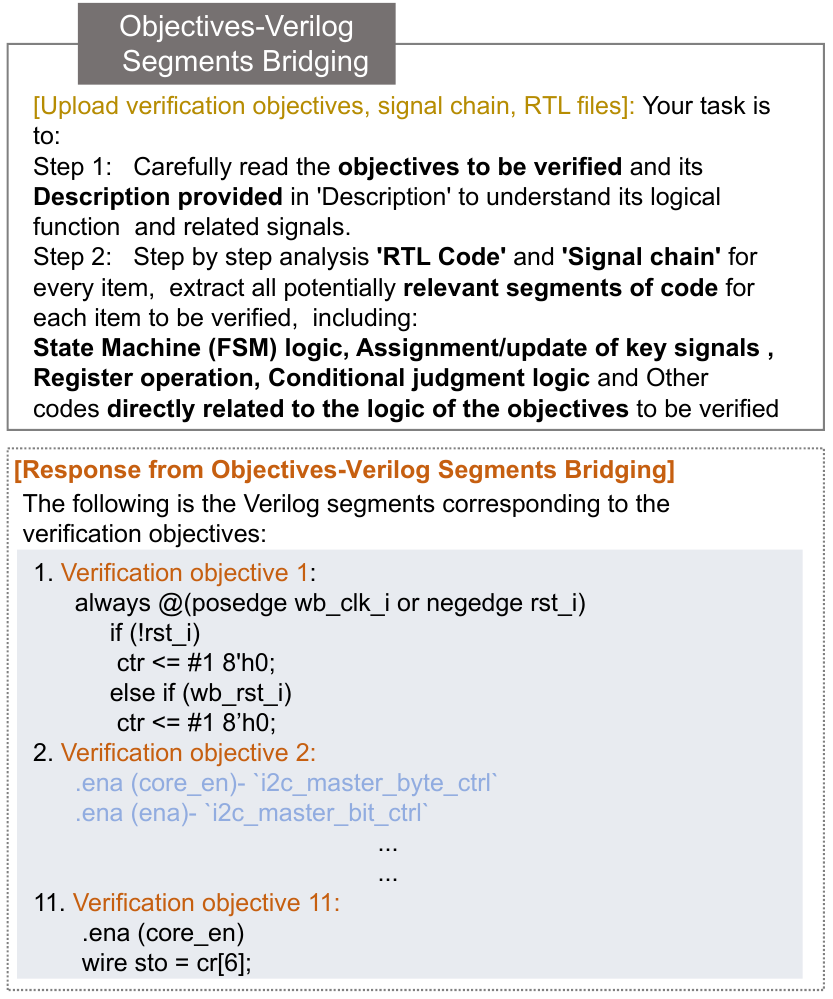}
\caption{Prompt and response of bridging verification objectives to verilog segments}
\label{Features-VerilogSnippetsBridging}
\end{figure}

\textbf{Bridging Verification Objectives to Verilog Segments:} After completing the extraction of the signal chain $\mathcal{SC}(s_{\text{0}})$, we input the verification objectives $\mathcal{V}$, signal chain $\mathcal{SC}$ and design file collection $\mathcal{D} = \{d_1, d_2, \dots, d_m\}$ into the LLM to perform document scanning and matching operations. We define the bridging function as:
\begin{equation}
    \mathcal{M} : (\mathcal{V}, \mathcal{SC}, \mathcal{D}) \rightarrow \mathcal{R}_\mathcal{V}
\end{equation}
where $\mathcal{R}_v \subseteq D$ denotes the set of verilog code segments semantically related to the verification objectives $\mathcal{V}$. The matching process compares the signal semantics in $v_i$ with the defined locations of each signal $s_i \in \mathcal{SC}$ in the design files, establishing a structural alignment between verification requirements and RTL designs, the prompt of this process is shown in Fig. \ref{Features-VerilogSnippetsBridging}.
\subsection{\textbf{Step 4: SVA Generation}}
In this stage, we used GPT-4o with verification objectives, verilog segments, and signal chain context to generate SVAs. However, initial evaluations revealed that despite correct intent, the outputs suffered from inconsistent formatting. These flaws made the code unreliable and required significant manual correction, indicating that the model needs fine-grained guidance to ensure structural consistency.

To address this issue, we design a fine-grained prompt strategy to guide the generation process and enforce format alignment. Specifically, for each signal or functional behavior to be verified, we incorporate the following constraints into the prompt:
\begin{enumerate}[]
    \item \textbf{Signal Binding:} Signals are explicitly referenced and bound using the format:
    \small{
    \begin{equation*}
    \begin{aligned}
    &\texttt{{source\_module\_name}.{signal\_name}}
    \end{aligned}
    \end{equation*}}
    \item \textbf{Unified Output Format:} All generated assertions follow a consistent template:
\begin{equation*}
\scriptsize
\begin{aligned}
&\texttt{assert property} \\
&\texttt{@(posedge \{source\_module\_name\}.\{signal\_name\})} \\
\end{aligned}
\end{equation*}
\end{enumerate}

By adopting this format, we effectively constrain the LLM's generation behavior to enhance the reliability of GPT-4o’s assertion outputs. 
\section{experiments}
\subsection{Experimental Setup}
The benchmarks employed in this study are derived from open-source datasets\cite{b18}\cite{b19}. The external inputs for the framework consist of initial specification documents in PDF format, signal definition files, and RTL designs in Verilog format.

To evaluate the effectiveness of AssertGen against SOTA approaches, three assertion generation frameworks—AssertLLM\cite{b10}, SPEC2Assertion\cite{b20}, and AssertGen—are compared, with their correctness and quality assessed using the commercial formal verification tool Cadence JasperGold (v21.12.002). Moreover, all experiments are run on a server with an Intel(R) Xeon(R) Gold 6148 CPU @ 2.40GHz.
\subsection{Evaluation Metrics}
Our evaluation protocol is designed for objectivity and rigor. All SVAs generated by the LLM are used directly without post-processing, and their quality is assessed using metrics from JasperGold, with the mutation testing (MT) as well. Unlike simpler structural metrics such as Cone of Influence (COI) or Proof Core (PC) provided by Jaspergold, MT directly evaluates an assertion's quality with its capability to detect injected mutation errors. It verifies that an SVA is not merely valid, but also capable of detecting injected design faults, making it the most stringent evaluation of an assertion’s verification power. In line with the methodology from\cite{b10}, our protocol for determining metrics like $FPR$ conservatively treated any proof that timed out after 5 hours as a pass. The detailed evaluation criteria are shown in Table \ref{Summary of Evaluation Metrics}.
\begin{table}[h]
\centering
\caption{Summary of Evaluation Metrics}
\label{Summary of Evaluation Metrics}
\begin{tabular}{|c|c|}
\hline
\textbf{Evaluation Metrics} & \textbf{Summary} \\ \hline
\makecell{\textbf{FPV pass rate ($FPR$)}} & \makecell{\hspace*{-50pt} Rate of SVAs passing FV} \\ \hline
\makecell{\textbf{Cone of influence ($COI$)}} & \makecell{\hspace*{-45pt} All logic driving a property} \\ \hline
\makecell{\textbf{Proof core ($PC$)}} & \makecell{\hspace*{-30pt} Minimal logic to prove property} \\ \hline
\makecell{\textbf{Bug Detection Rate ($BDR$)}} & \makecell{\hspace*{-6pt} Design error detection capability in MT} \\ \hline
\end{tabular}
\end{table}
\subsection{Benchmarks}
To evaluate the effectiveness and scalability of AssertGen, we select benchmark designs with varying complexities, primarily determined by the number of flip-flops (FFs) and primary inputs (bits), causing exponential growth in state and search spaces, respectively \cite{b21}. As detailed in Table \ref{Summary of Design}, our benchmark suite rigorously evaluates AssertGen by covering small-scale, control-intensive I\textsuperscript{2}C core (approximately 1282 lines of code (LOC)), as well as medium-scale designs like the compute-focused SHA3 and the size-optimized RISC-V CPU PicoRV32. To further test its capability, the suite includes the medium-to-large, mixed-signal ECG module. Finally, to push the limits of scalability, the suite is anchored by the large-scale Pairing cryptographic accelerator, which contains over 31,000 FFs. All post-synthesis metrics are obtained using the open-source tool Yosys-STA\cite{b22} under a 45nm process node, ensuring a consistent basis for our evaluation.
\begin{table}[h]
\centering
\caption{Benchmarks information}
\label{Summary of Design}
\begin{tabular}{|c|c|c|c|c|}
\hline
\rowcolor{gray!20}
\textbf{Design Name} & \textbf{Cell number}& \textbf{FF} & \textbf{Primary inputs (bits)}& \textbf{LOC}\\ \hline
\makecell{\textbf{I\textsuperscript{2}C}}  & \makecell{756}& \makecell{11}& \makecell{33}& \makecell{1282}\\ \hline
\makecell{\textbf{ECG}}  & \makecell{59084}& \makecell{7443}& \makecell{932}& \makecell{1635}\\ \hline
\makecell{\textbf{Pairing}}  & \makecell{228287}& \makecell{31380}& \makecell{1943}& \makecell{2145}\\ \hline
\makecell{\textbf{SHA3}}  & \makecell{22228}& \makecell{2211}& \makecell{585}& \makecell{618} \\ \hline
\makecell{\textbf{PicoRV32}}  & \makecell{17075}& \makecell{1969}& \makecell{2788}& \makecell{3049} \\ \hline
\end{tabular}
\end{table}

\subsection{Experimental Results}
To comprehensively evaluate the effectiveness of the AssertGen, we experimentally compare it with baseline methods on the five benchmark designs of different sizes. The detailed results are shown in Table \ref{Performance comparison of AssertLLM and AssertGen}.
\begin{table*}[t!]
\centering
\caption{Performance comparison of AssertLLM, SPEC2Assertion and AssertGen}
\label{Performance comparison of AssertLLM and AssertGen}
\begin{tabularx}{\textwidth}{
  >{\centering\arraybackslash}X | 
  *{5}{>{\centering\arraybackslash}X|} 
  *{5}{>{\centering\arraybackslash}X|} 
  *{4}{>{\centering\arraybackslash}X|} 
  >{\centering\arraybackslash}X 
}
\toprule
\multirow{2}{*}{Metric} & \multicolumn{5}{c|}{\cellcolor{gray!30}\textbf{AssertLLM}} & \multicolumn{5}{c|}{\cellcolor{gray!30}\textbf{SPEC2Assertion}}&\multicolumn{5}{c}{\cellcolor{mycyan!30}\textbf{AssertGen}} \\
\cline{2-16}
& \rule{0pt}{2.5ex}I\textsuperscript{2}C & ECG & \makebox[20pt][r]{Pairing} & \makebox[18pt][r]{SHA3} & \makebox[27pt][r]{PicoRV32} & I\textsuperscript{2}C & ECG & \makebox[20pt][r]{Pairing} & \makebox[18pt][r]{SHA3} & \makebox[27pt][r]{PicoRV32} & I\textsuperscript{2}C & ECG & \makebox[20pt][r]{Pairing} & \makebox[18pt][r]{SHA3} & \makebox[27pt][r]{PicoRV32} \\
\midrule
\makebox[22pt][r]{$FPR$(\%)} & 58.47 & 54.55 & 76.67 & 78.05 & 32.91 & 32.58 & 65.52 & 36.59 & 66.67 & 39.74 & \cellcolor{mycyan1!50}84.85 & \cellcolor{mycyan1!50}76.92 & 75.96 & \cellcolor{mycyan1!50}82.61 & \cellcolor{mycyan1!50}81.44 \\
\makebox[22pt][r]{$COI$(\%)} & 86.71 & 99.52 & 33.19 & 89.56 & 33.37 & 98.41 & 98.37 & 51.08 & 70.45 & 42.38 & \cellcolor{mycyan1!50}98.80 & \cellcolor{mycyan1!50}99.73 & \cellcolor{mycyan1!50}54.48 & \cellcolor{mycyan1!50}89.61 & \cellcolor{mycyan1!50}62.49\\
\makebox[22pt][r]{$PC$(\%)} & 4.43 & 0.22 & 0.05 & 58.90 & 12.18 & 0.26 & 3.58 & 0.08 & 0 & 13.27 & \cellcolor{mycyan1!50}98.35 & \cellcolor{mycyan1!50}23.55 & \cellcolor{mycyan1!50}5.32 & \cellcolor{mycyan1!50}60.79 & \cellcolor{mycyan1!50}32.21\\
\bottomrule
\end{tabularx}
\end{table*}
AssertGen delivers acceptable performance on smaller and medium-sized design modules. For the I\textsuperscript{2}C design, AssertGen achieves an FPR of 84.85\%, significantly outperforming both AssertLLM (58.47\%) and SPEC2Assertion (32.58\%). In terms of COI coverage, AssertGen provides a modest improvement over SPEC2Assertion (98.41\%) and AssertLLM (86.71\%). More importantly, for PC coverage, AssertGen reaches 98.35\%. This increase is mainly due to the I\textsuperscript{2}C module’s structure, where functionality centers around a few key registers, making high coverage on critical components more feasible. For the SHA3 module, AssertGen demonstrates equally comprehensive performance. It not only surpasses all baseline methods in FPR (82.61\%) and PC coverage (60.79\%), but also matches the top-performing baseline in COI coverage at 89.61\%, ensuring both the breadth and depth of verification. On the mid-scale PicoRV32 design, AssertGen still shows some advantages, achieving roughly twice the PC coverage of the comparative algorithms.

When applied to larger and more complex designs, AssertGen continues to demonstrate its effectiveness. For Pairing, AssertGen's FPR is 75.96\%, slightly lower than AssertLLM's 76.67\%. This difference may result from the complex temporal and logical relationships among cross-layer signals in Pairing, which make accurate assertion localization difficult. Regarding PC coverage, the baseline methods are nearly ineffective, yielding only 0.05\% and 0.08\% respectively, whereas AssertGen boosts this coverage to 5.32\%. On another large-scale design, ECG, AssertGen demonstrates improvements across all three metrics compared to the other two methods, achieving an FPR of 76.92\%, COI of 99.73\%, and PC of 23.55\%.

\subsection{Extended Experiment on Mutation Testing}
In the previous subsection, the effectiveness of the assertions generated by AssertGen is evaluated from the two dimensions of breadth and depth of structural coverage through the analysis of COI and PC. However, these two metrics cannot directly evaluate the ability of assertions to detect actual design errors. Therefore, in this subsection, we use mutation testing to evaluate the ability of generated assertions to detect mutation errors.

\begin{figure}
\centering
\includegraphics[width=70mm]{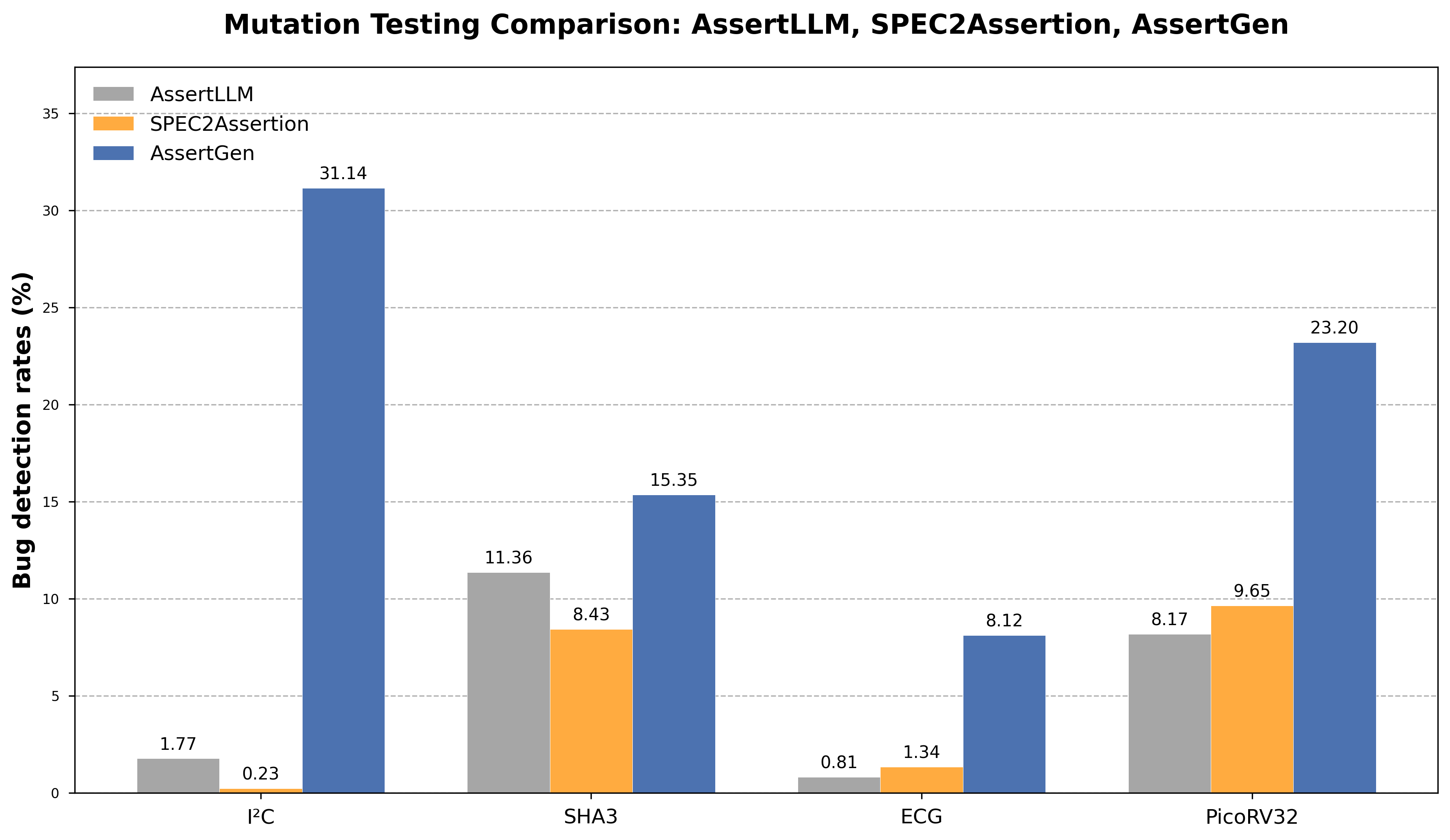}
\caption{Bug Detection Rates of AssertLLM, SPEC2Assertion, and AssertGen Across Four Designs}
\label{ab}
\end{figure}

The mutation testing results of AssertGen, Assertllm, and SPEC2Assertion are shown in Fig. \ref{ab}. AssertGen demonstrats certain bug detection capabilities on I\textsuperscript{2}C and SHA3 modules. For I\textsuperscript{2}C design, its BDR reaches 31.14\%, while the two baseline methods are both below 2\%, which makes it almost impossible to detect bugs. 
On the SHA3 module, AssertGen's detection rate reaches 15.35\%, which is nearly 4 percentage points higher than the best performing baseline method AssertLLM (11.36\%).

AssertGen still has some effect on the more complex designs of PicoRV32 and EGC. On the PicoRV32 core, its BDR (23.20\%) is twice as high as the baseline, which demonstrates that AssertGen has scalability. Meanwhile, on the ECG module, AssertGen achieves rates of 8.12\%, outperforming the baseline methods (with only 1.34\% and 0.81\% respectively). The largest module, Pairing, is excluded from this experimental evaluation due to its prohibitive computational cost, with an estimated runtime for mutation testing exceeding 120 hours. As SVAs are usually designed for coverage of certain important design properties, it is not necessary to has high detection rates of mutation test. But the improvement of BDR of AssertGen on the baseline methods still shows its capabilities on capturing bugs difficult to catch.

\section{Conclusion}
This paper presents AssertGen, a framework that automatically generates SVAs from natural language specifications and RTL design codes. Through a multi-stage modeling approach that aligns verification objectives, signal traces, and design codes, AssertGen outperforms SOTA methods like AssertLLM and SPEC2Assertion in several key metrics, showing improvements in terms of formal correctness (FPR), structural coverage (COI and proof core), and bug detection in mutation testing. Future work will explore and enhance the corner case bug detection capability of generated assertions.

\end{document}